\documentclass{ws-procs9x6}
\usepackage{latexsym}

\begin{document}
\title{DARK ELECTRIC MATTER OBJECTS: HISTORY OF DISCOVERY, MODES OF 
INTERACTION WITH MATTER, SOME INFERENCES AND PROSPECTS}

\author{E. M. DROBYSHEVSKI and M. E. DROBYSHEVSKI}

\address{Ioffe Physico-Technical Institute, Russian Academy of Sciences\\
St.Petersburg, 194021 Russia; E-mail: emdrob@mail.ioffe.ru}

\begin{abstract}
Experiments with thin ZnS(Ag) scintillators provide evidence with C.L. $>$ 
99.99{\%} for the existence of DArk Electric Matter Objects - daemons 
(presumably negatively charged Planckian particles with $M \sim $ 10$^{-5}$ 
g) captured from the Galactic disk into near-Earth, almost circular 
heliocentric orbits (NEACHOs). Their flux at $V \approx $ 10-15 km/s was 
found to be as high as $f_{\oplus } >$ 10$^{-7}$ cm$^{-2}$s$^{-1}$ and vary 
with $P$ = 0.5 y, with maxima in March and September. A daemon flux 
$f_{\oplus } \sim $ 10$^{-7}$-10$^{-6}$ cm$^{-2}$s$^{-1}$ is capable of 
accounting for the Troitsk anomaly in the $^{3}$T $\beta $-spectrum and 
suggests its more pronounced manifestation in future KATRIN experiment. In 
view of the channeling effect on iodine recoil nuclei in the NaI(Tl) 
crystal, the DAMA/NaI experiment is also apparently detecting a flux of 
daemons, $f_{\oplus } \sim $ 6$\times $10$^{-7}$ cm$^{-2}$s$^{-1}$, but in 
this case of those falling with $V$ = 30-50 km/s from strongly elongated, 
Earth-crossing heliocentric orbits (SEECHOs) oriented in the antapex 
direction, as a result of which the number of events detected in the 2-6-keV 
interval varies with $P$ = 1 y.
\end{abstract}

\bodymatter

\section{Planckian DM objects and possibility of their detection}
Our Universe started from Planckian scales, and it appears only reasonable 
to assume that the larger part of its mass, Dark Matter (DM), remains 
confined in Planckian objects, elementary black holes whose gravitational 
radius cannot be smaller than their Compton wavelength. Such objects with $M \sim $ 10$^{-5}$ g and $r_{g} \sim $ 10$^{-33}$ 
cm can carry an electric charge of up to $Z$e $\approx  G^{1/2}M \approx $ 
10e. While the properties of similar DArk Electric Matter Objects, 
\textit{daemons}, were considered by a number of authors [1-8], but the possibility of their 
detection was perceived skeptically (e.g., [1]).

We started from a notion that negative daemons are nuclear-active 
particles. Their capture of atomic nuclei should give rise to a 
release of energy $W \approx $ 1.8\textit{ZZ}$_{n}A^{-1/3}$ MeV ($\sim $10$^{2}$ MeV), 
with the attendant emission of scintillation-active electrons and nucleons 
[9]. Becoming confined in the remainder of the nucleus, the daemon should 
bring about successive decay of nucleons [10], thus lowering the charge of 
the remainder to $\vert Z_{n}\vert -\vert Z\vert  <$ 0, which 
makes it possible for the daemon to capture another nucleus with emission of 
new particles, etc. We noted also that a negative daemon is capable of 
catalyzing fusion of light nuclei, up to $Z_{n} \sim $ 6-9 (an analog of 
the muon catalysis of deuterons [11]). It should be pointed out that our 
understanding of the properties and behavior of daemons developed parallel 
to and in intimate connection with the experiments we were conducting.

Adopting $\sim $0.3 GeV/cm$^{3}$ for the DM density in the galactic 
halo, we arrive at $\sim $5$\times $10$^{-12}$ cm$^{-2}$s$^{-1}$ for the 
halo daemon flux intercepted by the Earth. The same applies to the galactic disk DM population with velocity dispersion of 4-30 km/s. 
The Sun moves relative to the disk objects with $V \approx $ 20 km/s (in the 
direction of the apex, which does not lie in the plane of ecliptic [12]). 
For such a velocity, the effective cross section of the Sun, amplified by 
gravitational focusing, exceeds its geometric cross section $\sim $10$^{3}$ 
times.

In crossing the Sun, daemons suffer slowing down, so that a sizable part of 
them falls back to move along strongly elongated and rapidly shrinking 
rosette-shaped trajectories, whose perihelia lie inside the Sun. These 
daemons settle eventually toward its center to form there a compact daemon 
kernel, which is capable of accounting for many specific features in solar 
physics, including generation of non-electron neutrinos [13]. If a daemon 
propagating along a rosette trajectory crosses the Earth's gravitational 
sphere of action, the perihelion of this trajectory has a high probability 
to leave the body of the Sun. This is how daemons enter and build up in 
stable SEECHOs (until their next encounter with the Earth). Straightforward 
estimates made in 1996 [14] in the gas-kinetic approximation of the mean 
free path suggest that the SEECHO daemon flux may reach as high as 
$f_{\oplus } \sim $ 3$\times $10$^{-7}$ cm$^{-2}$s$^{-1}$ for a velocity of 
30-50 km/s.

Recent celestial mechanics calculations [15] showed that SEECHOs crowd 
primarily in the ``shadow'' zone, i.e., the zone on the antapex side 
relative the Sun, because the petals of the rosette trajectories emerging 
into the apex hemisphere can no longer reach the Earth's orbit as a result 
of the slowing down the daemons undergo after several crossings of the Sun. 
Whence it follows that the flux of SEECHO daemons should exhibit a distinct 
one-year modulation with a maximum some time in June (it is appropriate to 
point out here that the direction to the apex is slightly different for 
different galactic disk populations, namely, stars of various classes, 
interstellar gas etc.; therefore, pinpointing this direction relative to the 
disk daemons is a task for the future).

Subsequent crossings by SEECHO objects of the Earth's gravitational sphere 
result in their gradual transfer to NEACHOs and accumulation there. It is 
from NEACHOs that daemons fall on the Earth with $V \approx $ 10(11.2)-15 
km/s.

\section{The history of the daemon discovery}
The above considerations, which were continually refined with due account of 
the building up experience, served as a basis for development of several 
generations of detectors.

In the very beginning (October 1996 - October 1998), we made an attempt at 
visualizing daemon-assisted catalysis of the fusion of light nuclei by 
detecting the sound wave generated by a large energy release along the 
trajectory (up to $\sim $10$^{3}$ erg/cm [14]). The experiments performed 
with lithium detectors demonstrated, however, extremely strong damping 
(along a distance of 2-3 cm) of sound with a characteristic frequency of 
$\sim $25 MHz. The experiments in which 45-mm-thick Be plates were used also 
did not produce promising results, the more so that the unavoidable increase 
here of the characteristic frequency up to $\sim $600 MHz would require 
development of unique methods of detection of such ultrasound. Therefore, 
both sides of the Be plates were coated with a $\sim $10-$\mu $m-thick layer 
of ZnS(Ag) scintillator, which could register jets of electrons and up to 
$\sim $10$^{4} \quad ^{18}$O nuclei (produced in fusion of Be nuclei) at the 
points of daemon entry into and exit out of the plates. Each side of the 
plate was viewed by a FEU-167 PM tube. Two such plates, each of 600 cm$^{2}$ 
area, were assembled into a telescope oriented continuously in the direction 
of the expected arrival of SEECHO daemons ($\sim $35$^{o}$ leading the Sun). 
A 700-h exposure accumulated in April-May 1999 did not yield the expected 
results; indeed, no scintillations with the shape characteristic of the 
passage of heavy particles (HPSs) and shifted by $\Delta t $ = 0.5-1.5 $\mu $s 
with respect to one another were observed [16]. An analysis of reasons 
for this failure led us to the understanding that the catalytic action of 
daemons should unavoidably be poisoned by their capture of heavy nuclei 
which are present in our beryllium produced by powder metallurgy ($\sim $0.1 
at.{\%} impurities, including Si, Fe etc.). This suggested immediately the 
possibility of daemon-stimulated nucleon decay [10].

It is this new ideology that underpinned the development in November 1999 of 
a four-module detector [17]. Each module consisted of a cubic iron-sheet 
container (0.3-mm-thick Fe sheet coated on both sides with 2 $\mu $m Sn 
layer), 51 cm on a side, covered on top with black paper. In the middle of the 
module, spaced by 7 cm, were fixed two transparent polystyrene plates, 4 
mm thick, separated by black paper. Their bottom faces were coated by a 
$\sim $3.6 mg/cm$^{2}$ ZnS(Ag) powder layer. Each plate was viewed on its 
side by a FEU-167 PM tube. We purposefully made the system sensitivity 
asymmetric for enhancing the differences in the up/down distributions 
in the number of time-shifted events. Initially (November-December 1999), 
all modules were also oriented in the direction of the expected arrival of 
SEECHO daemons from the Sun (including those crossing the Earth in the 
evening time).

%\begin{figure}[htbp]
%\centerline{\includegraphics[width=10.42in,height=8.33in]{Sydneytalk_drob21.eps}}
%\label{fig1}
%\end{figure}

\begin{figure}[t]
\begin{center}
\psfig{file=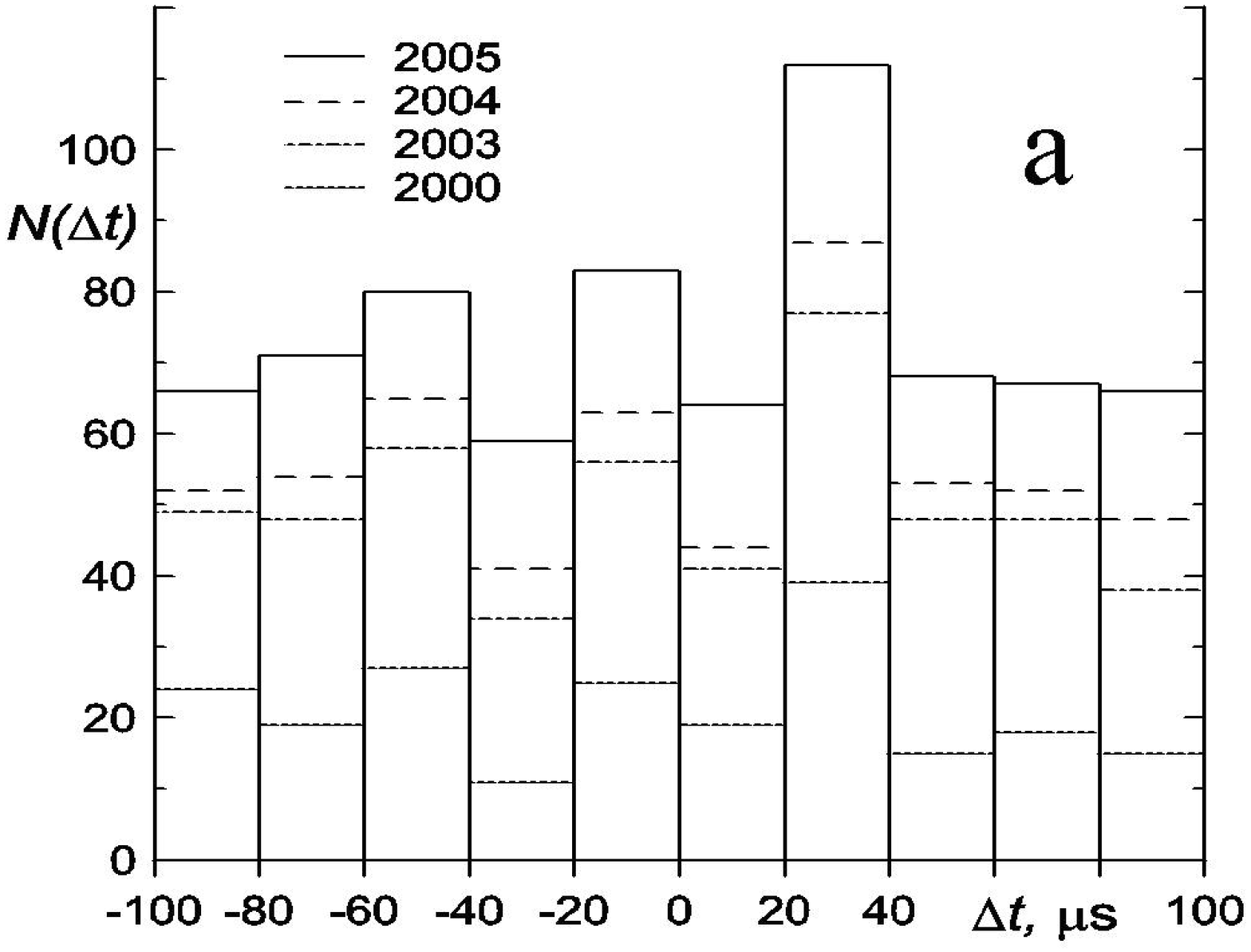,width=2.1in}
\psfig{file=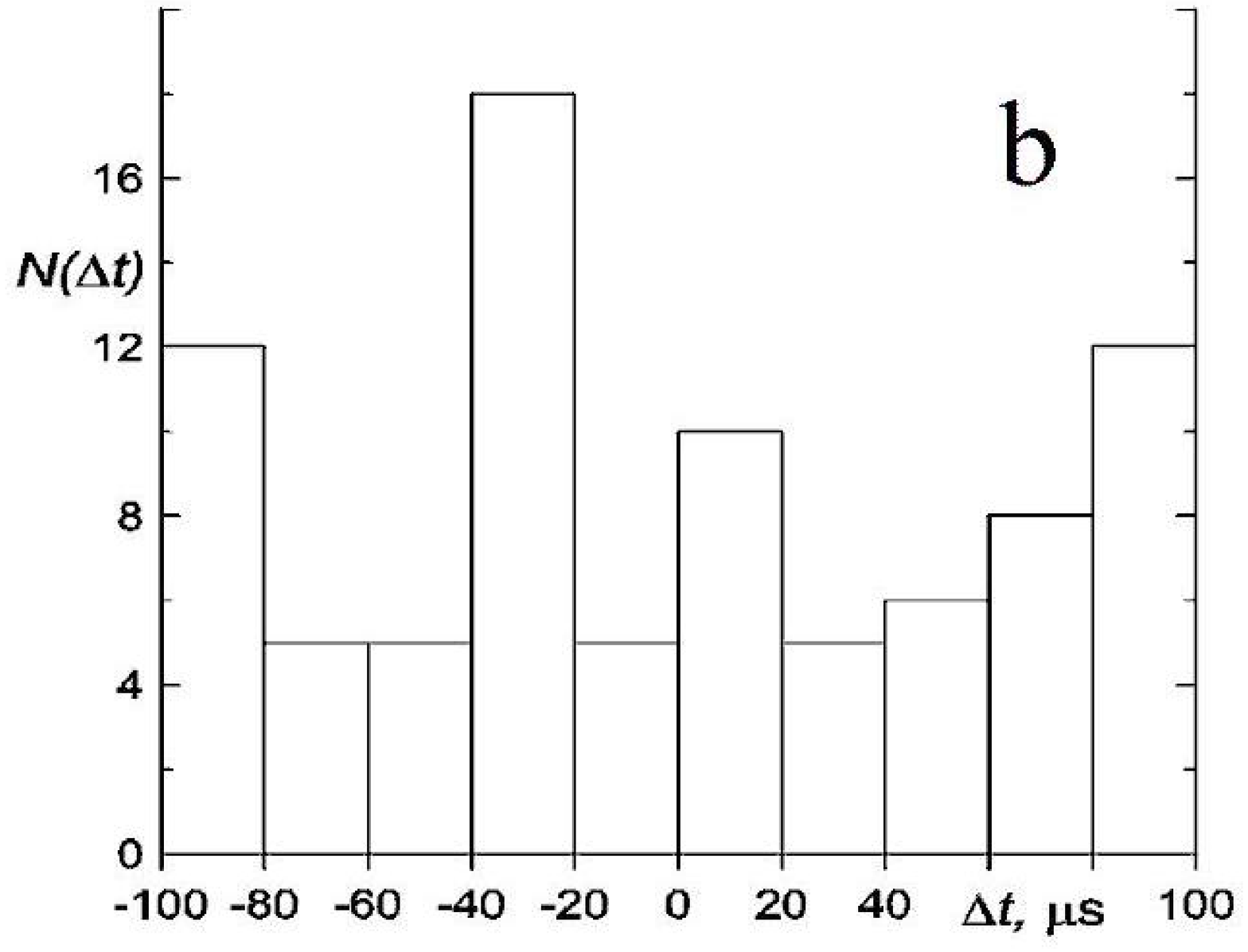,width=2.1in}
\end{center}
\caption{Distributions $N(\Delta t)$ of pair events on their time shift 
relative to the upper screen HPSs. (a) The sum of Marches 2000, 2003, 2004, 
and 2005. The +30$\mu $s maximum exceeds the mean level by 33 events; its 
significance is 3.63$\sigma $ (C.L. $>$ 99.97 {\%}) [18]. (b) The 
underground Baksan experiment [21]. The photo-cathode of the bottom PMT, 
which is sensitive to daemon passages due to its thick $\sim $1 $\mu $m 
inner Al coating, is screened with Al foil. Observations during September 
3-11, 2005; 86 events altogether. Significance of the --30 $\mu $s maximum 
is 2.2$\sigma $.}
%\label{aba:fig1}
\end{figure}

\subsection{Observation of NEACHO daemons in St. Petersburg}
In January 2000, we discontinued orienting the system; from that time on, 
the scintillators were always kept horizontally. In February, the 
$N(\Delta t)$ distribution revealed some features with a hint at statistical 
significance, so that while previously the detector was switched on for 
10-12 h in the daytime, starting with March 1, 2000, it was operated on the 
round-the-clock basis.

Figure 1a displays $N(\Delta t)$ distributions with HPSs on the top PM tube 
obtained in the March experiments in 2000, 2003, 2004, and 2005 [18]. The 
peak at 20 $< \Delta t <$ 40 $\mu $s stands out with a 99.97{\%} 
confidence. Accepting 29 cm as a base distance from the top ZnS(Ag) layer to 
the bottom lid of the container (see also Sec. 2.2 below), we arrive at a 
velocity of 10-15 km/s, a figure characteristic of objects falling from 
NEACHOs [17-19]. Assuming $\Delta $\textit{t $\approx $ }30 $\mu $s to be the time taken up by 
``digestion'' of a Zn nucleus captured in the ZnS(Ag) layer, we come to 
$\Delta $\textit{$\tau $}$_{ex} \sim $ 10$^{-6}$ s for the time of daemon-stimulated decay 
of a nucleon (proton or neutron) [19]. For such a value of $\Delta $\textit{$\tau $}$_{ex}$, 
the dimensions of our detector make it insensitive to the passage of daemons 
with $V >$ 30 km/s, even if they capture S nuclei in passing through the 
ZnS(Ag) layer.

\begin{figure}[t]
\begin{center}
\psfig{file=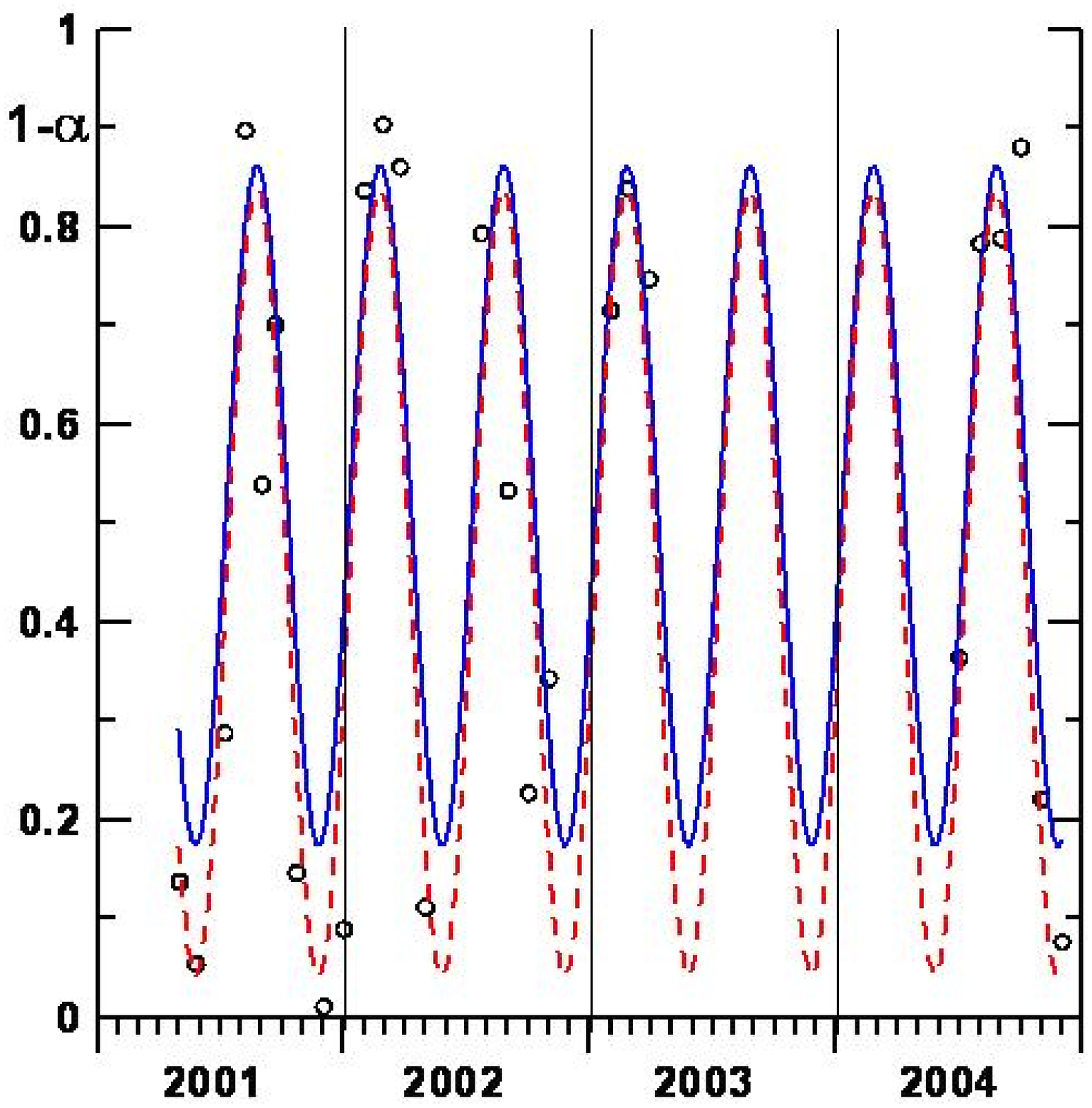,width=2.1in}
\psfig{file=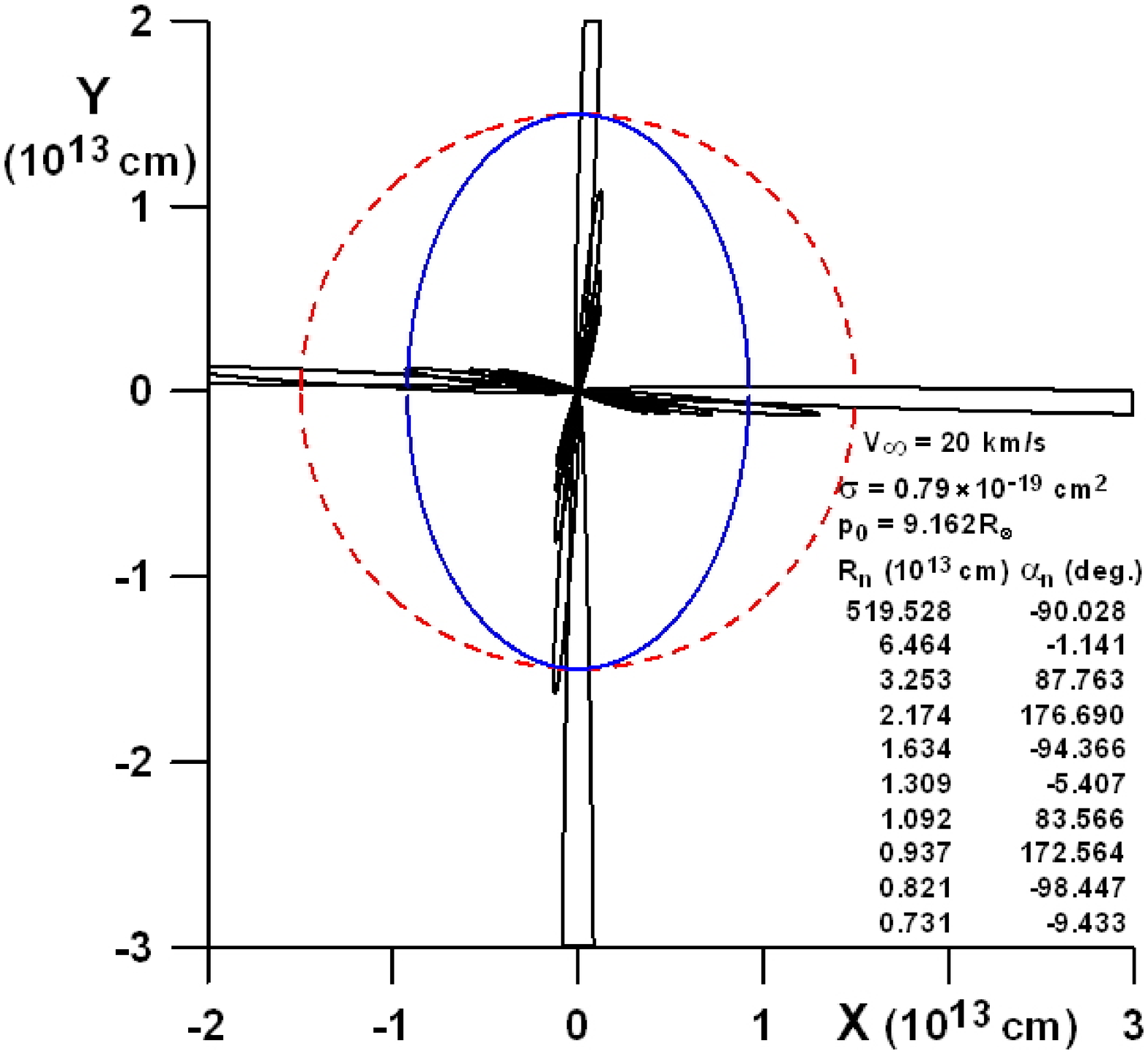,width=2.1in}
\end{center}
%\begin{tabular}{l}
\caption{Seasonal variation of 1 - \textit{$\alpha $}, the extent to which the distribution 
$N(\Delta t)$ (at --100 $< \Delta t <$ +100 $\mu $s) deviates from the 
constant level produced by background events.
(- - - -) Weights of all the points are equal, the correlation coefficient 
of the sine curve ($P$ = 0.5 yr) with the points is $r$ = 0.87, its C.L. $>$ 
99.9{\%}.
(------) Weights of the points are proportional to 1 - \textit{$\alpha $}; $r$ = 0.73, its C.L. = 99.3{\%}.} 

\caption{An example of multi-loop (cross-like) trajectory of an object 
being braked by the Solar matter at repeated passages through the Sun (its 
center is at X = 0, Y = 0) [15]. Object of 3$\times $10$^{-5}$ g mass and 
cross-section \textit{$\sigma $ }= 0.79$\times $10$^{-19}$ cm$^{2 }$falls from infinity (X = 
-$\infty $; $V_{\infty }$ = 20 km/s) with an impact parameter $p_{0}$ = 
Y(-$\infty )$ = 9.162R$_{\odot}$. The figure plane contains the apex direction 
and a normal to it lying in the ecliptic plane. An ellipse with the semi-major 
axis of 1 AU is the Earth's orbit projection; the dotted circle of 1 AU 
radius is given as a scale for the reader's orientation.}
%\end{tabular}

%\label{aba:fig1}
\end{figure}

After a full year of observations and testing of the equipment, we came to 
the conclusion that the NEACHO flux varies with $P$ = 0.5 y, with maxima in 
March and September (Fig. 2) [20]. This correlates well with our recent 
calculations of the capture of galactic disk daemons crossing the Sun. As 
the Sun moves in the apex direction relative to the daemon population of the 
galactic disk, daemons with an impact parameter of $\approx $9.2$R_{\odot}$ in 
the plane that contains the direction to the apex and is tilted at the 
minimum angle to the plane of ecliptic, cross the Sun with the corresponding 
slowing down and become captured into distinctive cross-shaped rosette 
trajectories (Fig. 3). Their petals, which are normal to the direction to 
the apex, cross \textit{repeatedly} the Earth's orbit close (\textit{sic}!) to the equinox zones, so that 
the daemons moving along them can eventually transfer to SEECHOs and, 
subsequently, to NEACHOs with a probability which is the highest compared 
with the other possible scenarios. It is these NEACHO daemons that are 
detected in March and September. Defining the force that slows down the 
daemon in the Sun by \textit{$\sigma \rho $V}$^{2}$ (\textit{$\rho $} is the density of solar material), observation 
of the maximum NEACHO daemon flux in these months (together with the maximum 
registration of SEECHOs by DAMA/NaI in June, see Sec. 4) yields for 
the daemon slowing-down cross section \textit{$\sigma $} $\approx $ 1$\times $10$^{-19}$ 
cm$^{2}$ [15].

\subsection{Underground experiments in Baksan with a daemon-sensitive PM tube}
Experiments performed with FEU-167 PM tubes (photocathode $\oslash $100 
mm, the bulb photocathode section $\oslash $125 mm) of different 
manufacturers (Svetlana in St. Petersburg and Ekran in Novosibirsk) showed 
that tubes with a thicker Al coating ($\sim $1 $\mu $m rather than $\sim 
$0.1 $\mu $m as specified) of the inner surface of the bulb photocathode 
section turn out to be sensitive to the passage of daemons with $V \sim $ 10 
km/s [18]. This is easy to understand because ($i)$ the path length of a 
complex with $Z$ = -1 to capture of an Al nucleus is $\sim $1 $\mu $m, and 
capture of a nucleus in the metal culminates in emission of many refilling 
electrons, and (\textit{ii}) passage of the complex, which consists of the daemon plus 
the remainder of the nucleus captured previously and now being digested, 
through 4-5 cm of vacuum inside the PM tube, is highly probable to give rise 
to the appearance and increase of the negative charge of the complex, 
because in vacuum this charge is not balanced by the positive charge of the 
nuclei captured for $Z$ = -1 in a path of 1-5 mm through air.

The module, in which only the top scintillator was left and the role of the 
bottom sensor was played by a FEU-167 with a thick inner Al coating whose 
photocathode was screened by aluminized Lavsan film, was first tried in the 
Baksan underground laboratory in September 2005 [21]. To reduce the Rn 
background, the detector was flushed by liquid N$_{2}$ vapor.

The experiment revealed a $\sim $hundredfold increase in detector 
sensitivity per unit area; indeed, the measured daemon flux with $V$ = 10-15 
km/s was as high as $\sim $10$^{-7}$ cm$^{-2}$s$^{-1}$, and it propagated 
from the interior of the Earth upward, with the maximum in the $N(\Delta 
t)$ distribution observed for -40 $< \Delta t <$ -20 $\mu $s (Fig. 1b). 
The underground measurements conducted in March 2006 (but without nitrogen 
flushing) revealed a standard peak at 20 $< \Delta t <$ 40 $\mu $s. This 
substantiated our choice of 29 cm between the top scintillator and the 
bottom lid of the module as the base distance (Sec. 2.1).

These results find a ready explanation if we assume that daemons build up in 
NEACHOs external to the Earth's orbit and, touching it on the outside, catch 
up with the Earth close to the equinoxes (actually, one to two weeks before 
that). This buildup in the external NEACHOs appears only natural, because in 
order for an object to be ejected out of the Earth's orbit its velocity has 
to be increased by $\Delta V \ge $ 12.3 km/s, whereas a single interaction 
with the Earth produces $\Delta V \le $ 11.2 km/s.

It thus follows that it is desirable to perform synchronous measurements in 
the Northern and Southern hemispheres; in the Southern hemisphere, in March 
the primary daemon flux will pass from bottom upward, and in September, from 
top downward.

\section{NEACHO {\&} GESCO daemons and the Troitsk anomaly in $^{3}$T $\beta 
$-spectrum}
Attempts at deriving the electron antineutrino mass from the end-point of 
the $^{3}$T $\beta $-spectrum ($E_{0}$ = 18574-18590 eV) revealed the 
presence near the end-point (at $E_{0}-E \approx $ 5 eV) of a step with an 
amplitude $\sim $1-3 mHz, which appears and varies slightly in position with 
$P$ = 0.5 y. This step was observed in the Troitsk experiment making use of an 
extended gaseous-tritium source of total horizontal area 0.35 m$^{2}$ (dia. 
5 cm, 3 m long plus $\sim $4 m-long channel leading to the spectrometer) 
[22]. No step is observed in Mainz experiments with a well devised solid 
tritium source.

A significant feature of the gas source, as we see it, is the use of superconducting windings of Nb, NbTi$_{2}$ and Nb$_{3}$Sn alloys 
needed to create the magnetic field channeling the $\beta $ electrons. 
Daemons crossing the windings capture Nb-containing atomic clusters and carry them across the gas source channel. Excitation of these clusters 
by nuclear-active daemons may conceivably bring about ejection of the inner 
$K$ electron of niobium (its binding energy $E_{k}$ = 18990 eV), with a 
resultant emission of niobium Auger electrons, in particular, of a very 
compact group of five lines (18566, 18568, 18569, 18570 and 18572 eV), which 
are close to $E_{0}$. The half-year periodicity and the phase of appearance 
of the step are clearly related with the variation of the low-energy NEACHO 
daemon flux (and the subsequent flux of objects captured from NEACHOs to 
GESCOs, geocentric Earth-surface crossing orbits). Adding the velocity of 
10-15 km/s to that of Auger electrons accounts for the dispersion in their 
energy $\Delta E \approx $ 5 eV [23].

Note that fine details in the emission of Nb $K$ electrons from a cluster 
containing Nb atoms remain unclear and require further consideration. The 
mechanism involved would be easier to understand if the daemon carrying 9, 
rather than 10, electronic charges would capture Sn nuclei in the windings 
($Z_{Sn}$ = 50). In these conditions, the positive charge of the daemon/Sn 
complex would be exactly equal to that of the Nb nucleus ($Z_{Nb}$ = 41). 
Shakeup of the electronic shells of such a complex, be it as a result of 
transitions of the Sn nucleus to lower-lying Rydberg levels in 
the daemon field (they lie within the electronic shells) or as a result of 
daemon-assisted catalysis of predominantly neutrons' decays in the Sn 
nucleus (see Sec. 5), would obviously give rise to emission of 
Auger electrons typical of Nb.

Even if only a third of the daemon-stimulated nucleon decays separated in 
time by $\sim $10$^{-6}$ s in the cluster nuclei are accompanied by emission 
of Auger electrons in the above-mentioned lines, the observed step amplitude 
should be reached for $f_{\oplus } \sim $ 10$^{-7}$-10$^{-6}$ 
cm$^{-2}$s$^{-1}$. It may be predicted that the KATRIN version with a 
gaseous T$_{2}$ source (and Nb-Sn containing windings) will reveal a still 
stronger ``Troitsk anomaly''.

\section{SEECHO daemons and the DAMA/NaI experiment}
There are grounds to believe that the one-year modulation of the number of 
scintillation signals at the 0.04 cpd/kg/keV level observed at a C.L. = 
6.3$\sigma $ in the DAMA/NaI experiment in Gran Sasso [24] can likewise be 
accounted for by daemons rather than by WIMPs as surmised by the authors.

This conclusion is argued persuasively for by the fact that the results 
obtained in DAMA/NaI are not reproduced in other experiments designed for 
detection of WIMPs with detectors of other types, although the sensitivity 
of the latter sometimes even exceeds noticeably that of the DAMA/NaI. To 
account for this, one is forced to devise rather exotic WIMPs [25].

Interestingly, the one-year modulation is observed only within a narrow 
range of scintillation amplitudes in NaI(Tl), which correspond to electron 
energies of 2-6 keV. But these are exactly the energies of the iodine 
recoils knocked out elastically by SEECHO daemons which fall on the Earth 
with $V$ = 30-50 km/s. The SEECHO flux reaches a maximum sometime in early June 
(see Sec. 1), which likewise conforms to the DAMA/NaI data.

At first glance, it might seem that the coincidence of the iodine recoil 
energies $E_{r}$ with the observed energy of scintillations produced by 
2-6-keV electrons is at odds with neutron beam calibrations which yield for 
$E_{r} >$ 10 keV of iodine a quenching factor $q <$ 10{\%} ($q$ is the ratio of 
scintillation signal intensities due to ions and electrons of equal 
energies). Taking into account ion channeling along some crystallographic 
directions changes the situation, however, because such ions impart nearly 
all of their energy to valence electrons of the crystal, so that for them 
$q \to $ 1. The channeling probability grows with decreasing $E_{r}$, so that 
for the iodine ion with $E_{r}$ = 4 keV emitted in an arbitrary direction it 
is \textit{$\eta $} $\approx $ 20{\%}. Thus, the efficiency of registration of 4-keV iodine 
ions by a NaI(Tl) crystal for $q$ = 1 should be $\approx $20{\%} [26].

Whence it follows that for producing the observed number of 2-6 keV 
events at the 0.04 cpd/kg/keV level in the 96-kg NaI(Tl) detector with an 
effective area of $\approx $1500 cm$^{2}$ one would have to assume a SEECHO 
daemon flux of about 0.04$\times $96$\times $4/86400$\times $1500$\times 
$\textit{$\eta $} = 6$\times $10$^{-7}$ cm$^{-2}$s$^{-1}$ [26]. Obviously enough, a number 
of attendant points still remain unclear, for instance, those associated 
with the mechanisms involved in the elastic knocking out of ions by daemons 
carrying remainders of previously captured nuclei, with a single-hit 
criterion application, the case where one takes into account events that 
happened in one out of the nine crystals in the DAMA/NaI experiment only, 
and so on. Let us hope that LIBRA experiment will provide information 
capable of shedding light on the situation.

\section{Attempts to reveal the daemon-stimulated proton decay}
Besides the existence itself of daemons, another basic assumption 
underpinning our experiment was gradual decomposition of a nucleus captured 
by the negative daemon residing inside it. We believed that the 
decomposition involved successive daemon-stimulated proton decays. The time 
taken up by a daemon-containing Zn nucleus to cross our detector ($\approx 
$30 $\mu $s) limits the mean proton decay time to $\Delta $\textit{$\tau $}$_{ex} \sim $ 1 
$\mu $s (Sec. 2.1).

To detect such a series of successive events, we assembled a detector of 
4.3-cm-thick CsI(Tl) crystals of 560 cm$^{2}$ total area. We hoped that for 
$f_{\oplus } \sim $ 10$^{-7}$ cm$^{-2}$s$^{-1}$ we would observe once every 
5 h a trail of scintillations spaced, on the average, by about $\sim $1 $\mu 
$s. Adopting a velocity $V$ = 10-15 km/s, we would observe 3 to 4 such 
scintillations.

Decay of a proton should release an energy of 938 MeV. If the proton 
decomposes into $\pi ^{+}$ and $\pi ^{0}$ mesons, they will leave only 
100 MeV in a 4.3-cm-thick CsI crystal.

However, at the threshold level of about 50 MeV, we did not observe triple 
events during weeks!

We thus come to a tentative conclusion that the daemon, a negatively charged 
Planckian black hole, decomposes the captured nucleus in a somewhat 
different way. It would be difficult to add something definite on this point 
at this time. One cannot exclude the possibility, for instance, that a 
nucleus decomposes by consecutive capture by the daemon of electrically 
neutral neutrons (by ``winding'' them on its gravitational horizon, as it 
were) accompanied by emission of non-detectable gravitons etc. In these 
conditions, the nucleus would preserve for a certain time a constant (the 
original) number of protons and its positive charge. The neutron-deficient 
nuclear remainder would then get rid of the large proton excess by rare 
(once every few tens of $\mu $s) fission events and/or practically 
simultaneous ejection of products of a new proton fusion ($\alpha $ 
particles and so on), a process similar to the one occurring in 
daemon-assisted catalysis of many ($\sim $10) protons in the Sun considered 
by us elsewhere [27]. The scenario of consecutive daemon-assisted decay of 
neutrons in a nucleus is argued for possibly by the multiple emission of 
Auger electrons by the daemon/captured Sn nucleus complex at $Z$ = 41 = 
\textit{constant} (see Sec. 3). The daemon itself also may free itself, but much more rarely, 
of the neutron matter wound on it by flash-evaporating it in the form of 
hard radiation, to regain its original Planckian mass.

\section{Concluding remarks and future prospects}
The daemon approach to the nature of DM which we are developing has turned 
out to be valid, self-consistent and fruitful. Fairly simple experiments not 
only substantiate the existence of daemons (by now, with a C.L. $>$ 
99.99{\%}) but suggest possible directions in which the daemon 
paradigm can be extended (the need of taking into account the Sun's motion 
relative to the galactic disk population, daemon-stimulated nucleon decay 
etc.).

It is essential that the existence of daemons and their buildup in the Solar 
system in Earth-crossing orbits of different types are argued for not only 
by our experiments but by the DAMA/NaI project as well. By the way, it 
becomes now clear why its results are not reproduced in other experiments 
aimed at detection of WIMPs, as well as that the unavoidable effect of 
channeling of low-energy iodine ions ($<$10 keV) should play a substantial 
part in interpretation of the DAMA/NaI results. All the features of the 
Troitsk anomaly in the tail of the $^{3}$T $\beta $-spectrum, whose 
existence was already being questioned by the very people who had discovered 
it because of its appearing to be impossible to explain, also find ready 
interpretation in terms of the daemon paradigm (we have in mind the 
amplitude and width of the step, the half-year periodicity and the phase of 
its appearance).

The agreement between the ground-level daemon fluxes estimated theoretically 
in 1996 ($\sim $3$\times $10$^{-7}$ cm$^{-2}$s$^{-1}$ [14]), measured by us 
experimentally in St. Petersburg and Baksan ($>$10$^{-7}$ cm$^{-2}$s$^{-1}$, 
2005 [18,21]), derived from the Troitsk anomaly (10$^{-7}$-10$^{-6}$ 
cm$^{-2}$s$^{-1}$, 2000 [23]), and following from DAMA/NaI studies ($\sim 
$6$\times $10$^{-7}$ cm$^{-2}$s$^{-1}$, 2003 [26]) is truly remarkable. 
Celestial mechanics calculations based on our and DAMA/NaI figures have 
yielded the cross section of daemon braking by solar matter of 
10$^{-19}$ cm$^{2}$.

Measurements of the direction of propagation of the primary daemon flux 
suggest daemon buildup in NEACHOs external to the Earth's orbit. It would 
thus be reasonable to carry out synchronous studies of the March and 
September fluxes in the Northern and Southern hemispheres. In the Northern 
hemisphere, in March, the primary daemon flux goes from above downward, 
while in September, as follows from the Baksan measurements, it passes 
upward, whereas in the Southern hemisphere the daemons should move in 
opposite directions relative to the observer.

The conclusions that can be drawn from the daemon paradigm concerning the 
existence of daemon kernels in the Earth and in the Sun permit us to 
understand or explain in a new way many geo- and heliophysical observations. 
The crowding of daemons toward the Galactic center clarifies the reason for 
the observed appearance of positrons there as a product of daemon-assisted 
catalysis of proton decay and/or fusion [27-28]. The same may apply to 
generation of the excess GeV radiation (see Sec. 5).

We have suddenly and completely unexpectedly found ourselves at the 
threshold of Planckian physics and painfully far away from an adequate 
understanding of the properties of daemons and of their interaction with 
matter. It appears therefore only natural that many of the above 
considerations may turn out in need of further refinement and 
critical analysis.
%\newline
\newline

\textbf{Acknowledgements}
\newline

EMD is greatly indebted to the Dark-2007 Organizers and to the Russian 
Foundation for Basic Research for financial support of his attending the 
Conference.\\

\textbf{References}
\begin{enumerate}
\item M.A.Markov, \textit{Sov. Phys. JETP}, \textbf{24}, 584-592 (1967).
\item K.P.Stanyukovich, \textit{Sov. Phys. Doklady}, \textbf{168}, 781-784 (1966).
\item S.Hawking, \textit{Mon. Not. Roy. Astron. Soc}., \textbf{152}, 75-78 (1971).
\item M.Turner, in ``\textit{Dark Matter in the Universe}'' (IAU Symp. No\textbf{117}), J.Kormendy {\&} G.R.Knapp (eds.), Dordrecht: Reidel, pp.445-488 (1987).
\item J.D.Barrow, E.J.Copeland, A.R.Liddle, \textit{Phys. Rev}., \textbf{D46}, 645-657 (1992).
\item P.Ivanov, P.Naselsky, I.Novikov, \textit{Phys. Rev}., \textbf{D50}, 7173-7178 (1994).
\item S.Alexeyev \textit{et al}., \textit{Class. Quantum Grav}., \textbf{19}, 4431-4443 (2002).
\item E.M.Prodanov, R.I.Ivanov, V.G.Gueorguiev, \textit{Astropart. Phys}., \textbf{27}, 150-154 (2007); \textit{hep-ph}/0703005.
\item E.M.Drobyshevski, \textit{Preprint PhTI}-\textbf{1663}, St-Petersburg (1996).
\item E.M.Drobyshevski, \textit{Mon. Not. Roy. Astron. Soc}., \textbf{311}, L1-L3 (2000).
\item S.S.Gershtein, L.I.Ponomarev, in ``\textit{Muon Physics}'', V.W.Hughes {\&} C.S.Wu (eds.), Acad. Press, v.\textbf{III}, pp.142-233 (1977).
\item C.W.Allen, \textit{Astrophysical Quantities}, 3$^{rd}$ ed., Univ. London, The Athlone Press (1973).
\item E.M.Drobyshevski, \textit{Astron. Astrophys. Trans}., \textbf{23}, 173-183 (2004); \textit{astro-ph}/0205353.
\item E.M.Drobyshevski, in ``\textit{Dark Matter in Astro- and Particle Physics (DARK'96)}'', H.V.Klapdor-Kleingrothaus {\&} Y.Ramachers (eds.), World. Sci., pp.417-424 (1997). 
\item E.M.Drobyshevski, M.E.Drobyshevski, \textit{Astron. Astrophys. Trans}., \textbf{26} (in press) (2007); \textit{arXiv}:0704.0982.
\item E.M.Drobyshevski, \textit{Physics of Atomic Nuclei}, \textbf{63}, 1112-1117 (2000).
\item E.M.Drobyshevski, \textit{Astron. Astrophys. Trans}., \textbf{21}, 65-73 (2002); see also \textit{astro-ph}/0007370$.$
\item E.M.Drobyshevski, \textit{Astron. Astrophys. Trans}., \textbf{25}, 43-55 (2006); \textit{astro-ph}/0605314.
\item E.M.Drobyshevski \textit{et al}., \textit{Astron. Astrophys. Trans}., \textbf{22}, 19-32 (2003); \textit{astro-ph}/0108231.
\item E.M.Drobyshevski \textit{et al}., \textit{Astron. Astrophys. Trans}., \textbf{22}, 263-271 (2003); \textit{astro-ph}/0305597.
\item E.M.Drobyshevski, M.E.Drobyshevski, \textit{Astron. Astrophys. Trans}., \textbf{25}, 57-73 (2006); \textit{astro-ph}/0607046.
\item V.M.Lobashev, \textit{Physics of Atomic Nuclei}, \textbf{63}, 1037-1043 (2000).
\item E.M.Drobyshevski, \textit{hep-ph}/0502056.
\item R.Bernabei \textit{et al}., \textit{Riv. Nuovo Cim.}, \textbf{20}, 1-73 (2003); \textit{astro-ph}/0307403.
\item R.Bernabei \textit{et al}., \textit{Int. J. Mod. Phys}., \textbf{A21}, 1445-1469 (2006).
\item E.M.Drobyshevski, \textit{arXiv}:0706.3095.
\item E.M.Drobyshevski, \textit{Mon. Not. Roy. Astron. Soc}., \textbf{282}, 211-218 (1996).
\item E.M.Drobyshevski, \textit{astro-ph}/0402367.
\end{enumerate}

\end{document}